\documentclass[12pt]{article}

\usepackage{amsmath}
\allowdisplaybreaks

\usepackage{amssymb}

\begin{document}

\baselineskip=16.8pt plus 0.2pt minus 0.1pt

\makeatletter
\@addtoreset{equation}{section}
\makeatother

\renewcommand{\theequation}{\thesection.\arabic{equation}}
\newcommand{\bm}[1]{\boldsymbol{#1}}
\newcommand{\calA}{{\mathcal A}}
\newcommand{\calB}{{\mathcal B}}
\newcommand{\calC}{{\mathcal C}}
\newcommand{\calE}{{\mathcal E}}
\newcommand{\calF}{{\mathcal F}}
\newcommand{\calG}{{\mathcal G}}
\newcommand{\calM}{{\mathcal M}}
\newcommand{\calN}{{\mathcal N}}
\newcommand{\calV}{{\mathcal V}}
\newcommand{\calK}{{\mathcal K}}
\newcommand{\calH}{{\mathcal H}}
\newcommand{\calI}{{\mathcal I}}
\newcommand{\calT}{{\mathcal T}}
\newcommand{\calU}{{\mathcal U}}
\newcommand{\calY}{{\mathcal Y}}
\newcommand{\calW}{{\mathcal W}}
\newcommand{\calL}{{\mathcal L}}
\newcommand{\calD}{{\mathcal D}}
\newcommand{\calO}{{\mathcal O}}
\newcommand{\calP}{{\mathcal P}}
\newcommand{\calQ}{{\mathcal Q}}
\newcommand{\calS}{{\mathcal S}}
\newcommand{\QB}{{\mathcal Q}_\text{B}}
\newcommand{\nn}{\nonumber}
\newcommand{\deriv}[2]{\frac{d #1}{d#2}}
\newcommand{\eps}{\varepsilon}
\newcommand{\veps}{\varepsilon}
\newcommand{\ds}{\displaystyle}
\newcommand{\Ke}{K_{\eps}}
\newcommand{\ce}{c_{\eps}}
\newcommand{\Psie}{\Psi_{\eps}}
\newcommand{\CR}[2]{\left[#1,#2\right]}
\newcommand{\ACR}[2]{\left\{#1,#2\right\}}
\newcommand{\tr}{\mathop{\rm tr}}
\newcommand{\Tr}{\mathop{\rm Tr}}
\newcommand{\p}{\partial}
\newcommand{\wh}[1]{\widehat{#1}}
\newcommand{\abs}[1]{\left| #1\right|}
\newcommand{\wt}[1]{\widetilde{#1}}
\newcommand{\KBc}{K\!Bc}
\newcommand{\VEV}[1]{\left\langle #1\right\rangle}
\newcommand{\Drv}[2]{\frac{\p #1}{\p #2}}
\renewcommand{\Re}{\mathop{\rm Re}}
\renewcommand{\Im}{\mathop{\rm Im}}
\newcommand{\EOMtest}{\text{EOM-test}}
\newcommand{\Anom}{A}
\newcommand{\eomT}{B}
\newcommand{\Ds}{\Delta_s}
\newcommand{\ee}{{\veps\eta}}
\newcommand{\etep}{{\eta\veps}}
\newcommand{\MK}{M_K}
\newcommand{\Dtau}{\Delta\tau}
\newcommand{\Mn}{EMNT}
\newcommand{\Mt}{\calM}


\begin{titlepage}

\title{
\hfill\parbox{3cm}{\normalsize KUNS-2415}\\[1cm]
{\Large\bf
Singularities in $\bm{K}$-space and Multi-brane Solutions\\
in Cubic String Field Theory
}}

\author{
Hiroyuki {\sc Hata}\footnote{
{\tt hata@gauge.scphys.kyoto-u.ac.jp}}
\ and
Toshiko {\sc Kojita}\footnote{
{\tt kojita@gauge.scphys.kyoto-u.ac.jp}}
\\[7mm]
{\it
Department of Physics, Kyoto University, Kyoto 606-8502, Japan
}
}

\date{{\normalsize September 2012}}
\maketitle

\begin{abstract}
\normalsize

In a previous paper [arXiv:1111.2389], we studied the multi-brane
solutions in cubic string field theory by focusing on the
topological nature of the ``winding number'' $\calN$ which counts the
number of branes.
We found that $\calN$ can be non-trivial owing to the singularity
from the zero-eigenvalue of $K$ of the $\KBc$ algebra, and that
solutions carrying integer $\calN$ and satisfying the EOM in the
strong sense is possible only for $\calN=0,\pm 1$.
In this paper, we extend the construction of multi-brane solutions
to $\abs{\calN}\ge 2$.
The solutions with $\calN=\pm 2$ is made possible by the fact that
the correlator is invariant under a transformation exchanging $K$
with $1/K$ and hence $K=\infty$ eigenvalue plays the same role as
$K=0$. We further propose a method of constructing solutions with
$\abs{\calN}\ge 3$ by expressing the eigenvalue space of $K$ as
a sum of intervals where the construction for $\abs{\calN}\le 2$ is
applicable.
\end{abstract}

\thispagestyle{empty}
\end{titlepage}


\section{Introduction}
\label{sec:Intro}

Construction of multi-brane classical solutions in cubic string
field theory (CSFT) \cite{CSFT} has attracted much attention
recently \cite{MS,Takahashi,HK,MS2,EM,AldoArroyo,MNT,BI}.
Most of the analytical solutions in CSFT are written in
pure-gauge form \cite{Okawa}, $\Psi=U\QB U^{-1}$, and let us consider
\begin{equation}
\calN=\frac{\pi^2}{3}\int\!\left(U\QB U^{-1}\right)^3 ,
\label{eq:calN}
\end{equation}
which is related to the energy density $\calE$ of the solution by
$\calE=\calN/(2\pi^2)$.\footnote{
We are considering static and translationally invariant solutions, and
have put both the space-time volume and the open string coupling
constant equal to $1$.}
Since the energy density of the tachyon vacuum is $-1/(2\pi^2)$,
the solution represents the $\calN+1$ branes.
An important property of $\calN$ is that it is a ``topological
quantity'' invariant under a small deformation of $U$.
The problem of constructing multi-brane solutions is equivalent
to understanding what kind of large deformation of $U$ can change
$\calN$.

In our previous paper \cite{HK}, we examined $\calN$ by focusing on
its similarity to its counterpart in the Chern-Simons (CS) theory in
three dimensions:
\begin{equation}
N_\text{CS}=\frac{1}{24\pi^2}\int_M\!\tr\left(gdg^{-1}\right)^3 .
\label{eq:N[g]}
\end{equation}
This is also a topological quantity and counts the winding number of
the mapping $g(x)$ from the manifold $M$ to the gauge group.
Then, a natural question is whether $\calN$ in CSFT also has an
interpretation as a kind of ``winding number'' taking possibly integer
values. For convenience, we call $\calN$ simply winding number in this
paper.

Our argument in \cite{HK} is restricted to the pure-gauge solution
of Okawa type \cite{Okawa} with $U$ given by\footnote{
In this paper, we adopt for convenience the non-Hermitian convention
for $\Psi=U\QB U^{-1}$. This can be converted to the Hermitian
one used in \cite{HK} by a gauge transformation.
}
\begin{equation}
U=1-Bc\left(1-G(K)\right) ,
\label{eq:U=1-FBcF}
\end{equation}
and specified by the function $G(K)$. Here, $(K,B,c)$ satisfies
the $\KBc$ algebra \cite{Okawa}:
\begin{equation}
\CR{B}{K}=0,\quad\left\{B,c\right\}=1,\quad
B^2=c^2=0,
\label{eq:CRalg}
\end{equation}
and
\begin{equation}
\QB B=K,\quad\QB K=0,\quad\QB c=cKc .
\label{eq:BRSTalg}
\end{equation}
For $U$ of \eqref{eq:U=1-FBcF}, $\Psi=U\QB U^{-1}$ is given by
\begin{equation}
\Psi=c\frac{K}{G(K)}Bc\left(1-G(K)\right) .
\label{eq:Psi=}
\end{equation}
In this paper, we assume that $G(K)$ is a rational function of $K$.

Let us recapitulate our results in \cite{HK}:
\begin{itemize}
\item
Corresponding to the fact that $N_\text{CS}$ \eqref{eq:N[g]} is
written as the integration of an exact quantity,
$N_\text{CS}=\int_M dH$, we showed that $\calN$ has the following
expression:
\begin{equation}
\calN=\int\!\QB\calA .
\label{eq:intQBA}
\end{equation}
In the case of $N_\text{CS}$, singularities of $H$ can make $N_\text{CS}$
non-vanishing; $N_\text{CS}=\int_{\p M}H\ne 0$ with $\p M$ being the
set of singularities of $H$.
Quite similarly, although the RHS of \eqref{eq:intQBA} vanishes
naively, it can take non-trivial values owing to the singularities
existent in $\calA$.\footnote{
Precisely speaking, these singularities mean those in the integrand of
the function $\calB(\tau)$ \eqref{eq:calB=intQB...} at $\tau=0$ or
$\tau=1$.}
We found that the origin of the singularity of $\calA$, and therefore
the origin of the winding number $\calN$ is the
{\em zero eigenvalue} of $K$.

\item
In order to properly evaluate the RHS of \eqref{eq:intQBA}, we
have to introduce a regularization for the singularity at $K=0$.
Since the eigenvalue distribution of $K$ is restricted to real and
non-negative, we adopted in \cite{HK} the $\Ke$-regularization of
making the following replacement:
\begin{equation}
K\to\Ke=K+\veps ,
\label{eq:KeReg}
\end{equation}
with $\veps$ being a positive infinitesimal.
In calculating the RHS of \eqref{eq:intQBA} in the
$\Ke$-regularization, we make the replacement \eqref{eq:KeReg}
after evaluating $\QB\calA$ by using \eqref{eq:BRSTalg}.
Then, we have
\begin{equation}
\int\!\left(\QB\calA\right)_{K\to\Ke}
\sim \veps\times\frac{1}{\veps}\ne 0 ,
\label{eq:ex(1/e)}
\end{equation}
where $\veps$ and $1/\veps$ are from the violation of the
BRST-exactness of the integrand due to the $\Ke$-regularization and
from the singularity of $\calA$ at $K=0$, respectively.

\item
We evaluated $\calN$ \eqref{eq:intQBA} for $G(K)$ of the form
\cite{ES,MS,MS2}
\begin{equation}
G(K)=\left(\frac{K}{1+K}\right)^n,
\qquad
(n=0,\pm 1,\pm 2,\cdots) .
\label{eq:G(K)=(K/(1+K))^n}
\end{equation}
We found that $\calN$ takes integer values, $\calN=-n$, only when
$n=0,\pm 1$. For $\abs{n}\ge 2$, $\calN$ deviates from the integer
$-n$; $\calN=\mp 2\pm 2\pi^2$ for $n=\pm 2$. For a generic $n$, we
obtain
\begin{equation}
\calN=-n+\Anom(n) ,
\label{eq:calN=-n+Anom}
\end{equation}
with
\begin{equation}
\Anom(n)=\frac{\pi^2}{3}\,n(n^2-1)\Re {}_1F_1(2+n,4;2\pi i) ,
\label{eq:Anom}
\end{equation}
where ${}_1F_1(\alpha,\gamma;z)$ is the confluent hypergeometric
function (see \eqref{eq:1F1}).\footnote{
$\Anom(n)$ is an odd function of $n$.
The same anomaly as \eqref{eq:Anom} appears in \cite{MS2}.
}
The result \eqref{eq:calN=-n+Anom}, whose derivation is
outlined in Appendix \ref{app:calNandEOMtest}, has been obtained by
expanding $G(K)$ around $K=0$; only the leading power $n$ of $G(K)$
determines $\calN$.
This is validated by the observation mentioned above that only
the singularity of $\calA$ at $K=0$ determines the value of $\calN$.
The anomaly $\Anom(n)$ in $\calN$ (i.e., the deviation of $\calN$ from
the integer $-n$) is inevitable except when $G(K)$ is finite or has
simple zero or simple pole at $K=0$. (As we shall see later, the
behavior of $G(K)$ at $K=\infty$ also affects the winding number. Here
and in \cite{HK}, we are assuming that $G(K=\infty)$ is finite and
non-vanishing.)

Here, we comment on the relation between our result
\eqref{eq:calN=-n+Anom} and that in the pioneering works
\cite{MS,MS2}. In \cite{MS,MS2}, they claim that $G(K)$ of
\eqref{eq:G(K)=(K/(1+K))^n} gives $\calN=-n$ without the anomaly
term for any $n$.
However, this is owing to their choice of deforming the contour
of $z$-integration along the pure-imaginary axis in the formula
\eqref{eq:szint} to the left $(\Re z<0$) for avoiding the singularity
at $z=0$. This manifestly contradicts with our $\Ke$-regularization
\eqref{eq:KeReg} which corresponds to integrating along
$\Re z=\veps>0$ and multiplying the integrand of \eqref{eq:szint} by
$e^{-\veps s}$.
The contour of \cite{MS,MS2} rather corresponds to $K\to K-\veps$,
which cannot work as a regularization.
In this paper, we admit that the anomaly in $\calN$ is unavoidable for
$\abs{n}\ge 2$ and pursue new ways of constructing solutions without
anomaly.

\item
The pure-gauge solution with the $\Ke$-regularization,
$\Psie=\left(U\QB U^{-1}\right)_{K\to\Ke}$, is no longer a pure-gauge,
and breaks the EOM by apparently $O(\veps)$.
We examined the EOM in the strong sense for $\Psie$:
\begin{equation}
\EOMtest[\Psi]=\int\!\Psie*\left(\QB\Psie+\Psie *\Psie\right)
=\veps\times\calE_\veps ,
\label{eq:EOMtest}
\end{equation}
with $\calE_\veps$ given by
\begin{equation}
\calE_\veps[G(K)]
=\int\!BcG(\Ke)c\frac{\Ke}{G(\Ke)}cG(\Ke)c\frac{\Ke}{G(\Ke)} .
\label{eq:calE_e}
\end{equation}
Since \eqref{eq:EOMtest} is multiplied by $\veps$, it vanishes unless
$\calE_\veps$ contains the $1/\veps$ singularity which again comes
from the zero eigenvalue of $K$. The situation is quite similar to the
case of the winding number $\calN$ \eqref{eq:intQBA}, and only the
leading power of $G(K)$ around $K=0$ determines $\EOMtest[\Psi]$
\eqref{eq:EOMtest}.
For $G(K)$ of \eqref{eq:G(K)=(K/(1+K))^n}, we found that
the EOM in the strong sense holds for $n=0,\pm 1$, but it is violated
for $\abs{n}\ge 2$. Concretely, we obtain (see Appendix
\ref{app:calNandEOMtest})
\begin{equation}
\EOMtest[\Psi]=\eomT(n)\equiv
\frac{n(n-1)}{\pi}\,\Im{}_1F_1(1+n,2,2\pi i),
\label{eq:EOM1F1}
\end{equation}
which vanishes only when $n=0,\pm 1$.\footnote{
$\eomT(n)$ has the property
$\eomT(-n)=(1+n)/(1-n) \eomT(n)$.
}
\end{itemize}

Summarizing, the singularity due to the zero eigenvalue of $K$ is the
origin of both the winding number $\calN$ and the $\EOMtest$.
The winding number $\calN$ deviates from integer and the EOM in the
strong sense is violated unless $G(K)$ is finite or has simple pole
or simple zero at $K=0$.
Namely, we can construct satisfactory (multi-)brane
solutions\footnote{
In this paper, we mean by a ``satisfactory solution'' a one
carrying an integer $\calN$ and satisfying the EOM in the strong
sense.
It is of course a problem in what sense the EOM should hold and
whether the EOM in the strong sense is sufficient.
We discuss this matter in Sec.\ 4.
}
only for the tachyon vacuum ($\calN=-1$), a single brane ($\calN=0$)
and two branes ($\calN=1$).

In this paper, we present a way to construct multi-brane solutions with
$\abs{\calN}\ge 2$. The point is that the singularities from
the $K=\infty$ eigenvalue as well as that from $K=0$ can be the
origins of the winding number $\calN$ \eqref{eq:intQBA}.
This is owing to the invariance of the correlator under a
transformation which replaces $K$ with $1/K$. This remarkable property
implies, in particular, that $G(K)$ and $G(1/K)$ give the same values
of $\calN$ and $\EOMtest$. By using the singularities from both $K=0$
and $K=\infty$, we can construct solutions with $\calN=\pm 2$ and
satisfying the EOM in the strong sense.
The corresponding $G(K)$ with $\calN=2$, for example, should behave
near $K=0$ and $K=\infty$ as $G(K)\sim 1/K$ ($K\to 0$) and
$G(K)\sim K$ ($K\to\infty$), respectively.
The singularity at $K=\infty$ needs a regularization besides the
$\Ke$-regularization \eqref{eq:KeReg} for $K=0$. We devise such a
regularization from the $\Ke$-regularization by using the
transformation $K\mapsto 1/K$ mentioned above.

However, it is impossible to construct satisfactory solutions with
$\abs{\calN}\ge 3$ since higher order zeros/poles of $G(K)$ at $K=0$
and $K=\infty$ inevitably lead to the anomaly in $\calN$ and the
breaking of the EOM in the strong sense.
As a possible resolution to this problem of constructing satisfactory
solutions with $\abs{\calN}\ge 3$, we consider using $G(K)$ which has
zeros/poles in $0<K<\infty$. Such $G(K)$ is ``dangerous'' since
its $\calN$ and $\EOMtest$ contain quantities which do not have
well-defined Schwinger parameter representations. We propose a way of
defining $\calN$ and $\EOMtest$ for such $G(K)$ by giving them as
a sum of well-defined ones obtained by expressing the eigenvalue space
of $K$, $[0,\infty]$, as a sum of intervals.
This division into intervals is analogous to the division of a sphere
$S^2$ into two hemispheres for defining the vector field of $U(1)$
monopole. By this method, we can construct satisfactory multi-brane
solutions carrying any integer $\calN$.

The organization of the rest of this paper is as follows.
In Sec.\ 2, we show that $K=\infty$ is equivalent to $K=0$, and can be
the origin of the winding number. This leads us to the construction of
satisfactory solutions with $\calN=\pm 2$.
In Sec.\ 3, we propose a way of defining solution carrying
$\abs{\calN}\ge 3$ by division of the eigenvalue space of $K$ into
intervals.
We summarize the paper and discuss future problems in Sec.\ 4.
In Appendix \ref{app:calNandEOMtest}, we present derivations of
eqs.\ \eqref{eq:calN=-n+Anom} and \eqref{eq:EOM1F1}.
In Appendix \ref{app:Proof_invBcccc}, we give a proof of the
invariance of the correlator under the transformation which exchanges
$K$ with $1/K$.

\section{$\bm{K=\infty}$ as another origin of $\bm{\calN}$}
\label{sec:Kinfty}

The winding number $\calN$ \eqref{eq:calN} in CSFT has another
expression \eqref{eq:intQBA}, which is the integration of a BRST-exact
quantity $\QB\calA$ and hence is apparently equal to zero.
We saw in \cite{HK} that $\calN$ can take non-trivial values due to
the singularity of $\calA$ at $K=0$. This allowed us to construct
satisfactory multi-brane solutions only for $\calN=0$ and $\pm 1$.
In this section, we show that another eigenvalue of $K$, $K=\infty$,
can play the same role as $K=0$ and hence can make $\calN$
non-trivial. By using the singularities both at $K=0$ and $K=\infty$,
we can extend the construction of satisfactory multi-brane solutions
to the cases $\calN=\pm 2$.

\subsection{Equivalence of $\bm{K=0}$ and $\bm{K=\infty}$}
\label{subsec:Equivalence}

We wish to show that $K=\infty$ is in a sense equivalent to $K=0$.
This is due to the invariance of the correlator on the infinite
cylinder under the following transformation which replaces $K$ with
$1/K$ and keeps the $\KBc$ algebra \eqref{eq:CRalg} and
\eqref{eq:BRSTalg}:
\begin{equation}
K\mapsto\wt{K}=\frac{1}{K},\qquad
B\mapsto\wt{B}=\frac{B}{K^2},\qquad
c\mapsto\wt{c}=cK^2Bc .
\label{eq:Kto1/K}
\end{equation}
Namely, the following surprising equality holds for any $\calW(K,B,c)$
with ghost number $3$:
\begin{equation}
\int\!\calW(K,B,c)
=\int\!\calW(\wt{K},\wt{B},\wt{c}) .
\label{eq:invBcccc}
\end{equation}
A proof of \eqref{eq:invBcccc} is given in
Appendix \ref{app:Proof_invBcccc}. There, we show \eqref{eq:invBcccc}
by using the $(s,z)$-integration expression of the correlator given in
\cite{MS,MS2}, and making a suitable change of integration variables.
We call the transformation \eqref{eq:Kto1/K} ``inversion'' in the rest
of this paper.

The inversion \eqref{eq:Kto1/K} is a special case ($g(K)=1/K$) of
a more general transformation $\Mt_g$ which keeps the $\KBc$ algebra
and is specified by a function $g(K)$ \cite{Erler,MNT,ErlerProc}:
\begin{equation}
\Mt_g(K)= g(K),\qquad
\Mt_g(B)=\frac{g(K)}{K}\,B,\qquad
\Mt_g(c)=c\frac{K}{g(K)}Bc .
\label{eq:Mtransf}
\end{equation}
We call the transformation $\Mt_g$ \eqref{eq:Mtransf}
Erler-Masuda-Noumi-Takahashi (\Mn) transformation hereafter.
Note that the correlator is not invariant under the \Mn\
transformation of an arbitrary $g(K)$. This is because the correlator
is not determined by the $\KBc$ algebra alone.
The inversion (the tilde operation) \eqref{eq:Kto1/K} is $\Mt_{1/K}$.
Note that $Bc$ is invariant under any \Mn\ transformation,
$\Mt_g(Bc)=Bc$.

The effect of the inversion \eqref{eq:Kto1/K} on
$\Psi$ of \eqref{eq:Psi=} is simply to replace $G(K)$ with $G(1/K)$.
This fact together with the invariance \eqref{eq:invBcccc} implies
that $G(K)$ and $G(1/K)$ should give the same $\calN$.
In this sense, $K=\infty$ is ``equivalent'' to $K=0$, and can be the
origin of the winding number and that of the breaking of the EOM in
the strong sense.
Recall that the formula \eqref{eq:calN=-n+Anom} counts the winding
number from the singularity at $K=0$ alone. Therefore,
the contribution of the singularity at $K=\infty$ to $\calN$ is
given by applying \eqref{eq:calN=-n+Anom} to $G(1/K)$.
Suppose that the leading behaviors of $G(K)$ near $K=0$ and $K=\infty$
are
\begin{equation}
G(K)\sim
\begin{cases}
K^{n_0} & (K\to 0)
\\
(1/K)^{n_\infty} & (K\to\infty)
\end{cases},
\label{eq:G(K)atK=0K=infty}
\end{equation}
with $n_0$ and $n_\infty$ being integers.
Then, the total winding number is given by
\begin{equation}
\calN=-n_0-n_\infty+\Anom(n_0)+\Anom(n_\infty) .
\label{eq:calN=-n_0+n_infty+A}
\end{equation}

\subsection{Regularization for both $\bm{K=0}$ and $\bm{K=\infty}$ }
\label{subsec:Reg}

For precisely defining $\calN$ and $\EOMtest$, we need a
regularization for $K=\infty$ as well as for $K=0$.
We introduce the regularization for $K=\infty$ as the map of the
$\Ke$-regularization \eqref{eq:KeReg} by the inversion
\eqref{eq:Kto1/K}.
Consider $(K,B,c)$ and $(\wt{K},\wt{B},\wt{c})$ related by
\eqref{eq:Kto1/K}. The regularization for $K=\infty$ is naturally
defined as the operation on $(K,B,c)$ induced by the regularization
for $\wt{K}=0$;
$(\wt{K},\wt{B},\wt{c})\to(\wt{K}+\eta,\wt{B},\wt{c})$ with $\eta$
being positive infinitesimal:
\begin{align}
K&=\frac{1}{\wt{K}}\to\frac{1}{\wt{K}+\eta}=\frac{K}{1+\eta K} ,
\nn\\
B&=\frac{\wt{B}}{\wt{K}^2}
\to\frac{\wt{B}}{\bigl(\wt{K}+\eta\bigr)^2}
=\frac{B}{\bigl(1+\eta K\bigr)^2} ,
\nn\\
c&=\wt{c}\wt{K}^2\wt{B}\wt{c}
\to \wt{c}\bigl(\wt{K}+\eta\bigr)^2\wt{B}\wt{c}
=c\bigl(1+\eta K\bigr)^2 Bc .
\label{eq:etaReg}
\end{align}
The regularization for $K$ is simply $1/K\to1/K+\eta$,
and we have to change $B$ and $c$ as well.
Let us introduce, in addition, the regularization for $K=0$. There are
two ways for this; one is to replace $K$ on the RHS of
\eqref{eq:etaReg} by $\Ke=K+\veps$, and the other is to start with
$(\Ke,B,c)$ in \eqref{eq:etaReg}. The regularized $(K,B,c)$ in each
case is then given by
\begin{equation}
K_\ee=\frac{\Ke}{1+\eta\Ke},\qquad
B_\ee=\frac{B}{\bigl(1+\eta\Ke\bigr)^2},\qquad
c_\ee=c\bigl(1+\eta\Ke\bigr)^2 Bc ,
\label{eq:KBc_ee}
\end{equation}
and
\begin{equation}
K'_\ee=\frac{K}{1+\eta K}+\veps,\qquad
B'_\ee=\frac{B}{\bigl(1+\eta K\bigr)^2},\qquad
c'_\ee=c\bigl(1+\eta K\bigr)^2 Bc .
\label{eq:KBc'_ee}
\end{equation}
These two way of regularization are interchanged by the inversion
\eqref{eq:Kto1/K}. Namely, the regularization by \eqref{eq:KBc_ee},
$(K,B,c)\to(K_\ee,B_\ee,c_\ee)$, is expressed in terms the tilded
variables of \eqref{eq:Kto1/K} by
\begin{equation}
(\wt{K},\wt{B},\wt{c})\to
\left(\frac{\wt{K}}{1+\veps\wt{K}}+\eta,
\frac{\wt{B}}{\bigl(1+\veps\wt{K}\bigr)^2},
\wt{c}\bigl(1+\veps\wt{K}\bigr)^2\wt{B}\wt{c}\right).
\end{equation}
In the rest of this paper, we mainly use the regularization by
\eqref{eq:KBc_ee}.

An important property of the regularized $(K,B,c)$, \eqref{eq:KBc_ee}
and \eqref{eq:KBc'_ee}, is that they satisfy the (anti-)commutation
relations of \eqref{eq:CRalg}:
\begin{equation}
\CR{B_\ee}{K_\ee}=0,\qquad
\ACR{B_\ee}{c_\ee}=1,\qquad
B_\ee^2=c_\ee^2=0 ,
\end{equation}
and the same equations for \eqref{eq:KBc'_ee}.\footnote{
More generally,
$$
K_{(g,h)}=g(K),\quad
B_{(g,h)}=\frac{h(K)}{K}\,B, \quad
c_{(g,h)}=c\frac{K}{h(K)}\,Bc,
$$
defined by $g(K)$ and $h(K)$ satisfy \eqref{eq:CRalg}.
They break the BRST algebra \eqref{eq:BRSTalg} unless $g(K)=h(K)$,
namely, unless they are a \Mn-transform of $(K,B,c)$.
}
Owing to this property, the present regularization is consistently and
unambiguously defined. Namely, the regularized quantities are
independent of whether we rewrite/simplify them by using the
(anti-)commutation relation \eqref{eq:CRalg} before or after the
replacement $(K,B,c)\to(K_\ee,B_\ee,c_\ee)$.
On the other hand, the BRST algebra \eqref{eq:BRSTalg} is broken by
the regularized $(K,B,c)$, \eqref{eq:KBc_ee} and \eqref{eq:KBc'_ee}.
This is necessary for making non-trivial the winding number
\eqref{eq:intQBA} given as the integration of $\QB\calA$ (see
\eqref{eq:ex(1/e)}).

The regularized version of the identity \eqref{eq:invBcccc} reads
\begin{equation}
\int\!\calW(K_\ee,B_\ee,c_\ee)
=\int\!\calW(\wt{K_\ee},\wt{B_\ee},\wt{c_\ee}) ,
\label{eq:ReginvBcccc}
\end{equation}
where $\wt{K_\ee}$, for example, is the inversion of $K_\ee$.
Explicitly, we have
\begin{equation}
\wt{K_\ee}=\frac{1}{K'_{\etep}},\qquad
\wt{B_\ee}=\frac{B'_\etep}{(K'_\etep)^2},\qquad
\wt{c_\ee}=c'_\etep(K'_\etep)^2B'_\etep c'_\etep ,
\label{eq:wtKBC_ee}
\end{equation}
which also satisfy the (anti-)commutation relations \eqref{eq:CRalg}.

\subsection{EOM in the strong sense}
\label{subsec:EOM}

Let us consider $\Psi$ \eqref{eq:Psi=} in the regularization of
\eqref{eq:KBc_ee}:
\begin{equation}
\Psi_\ee=\Psi\bigr|_{(K,B,c)\to(K_\ee,B_\ee,c_\ee)}
=c\frac{\Ke^2}{K_\ee G(K_\ee)}Bc\bigl(1-G(K_\ee)\bigr) .
\label{eq:Psi_ee}
\end{equation}
Due to the regularization, $\Psi_\ee$ no longer satisfies the EOM
exactly, and the breaking consists of two terms which are apparently
of $O(\veps)$ and $O(\eta)$, respectively.
Correspondingly, the EOM in the strong sense is given by
\begin{equation}
\EOMtest[\Psi]=\int\Psi_\ee*\left(
\QB\Psi_\ee+\Psi_\ee *\Psi_\ee\right)
=\veps\times\calE_\ee +\eta\times\calF_\ee ,
\label{eq:Eomtest=eE+etF}
\end{equation}
where $\calE_\ee$ and $\calF_\ee$ are expressed in terms of $K_\ee$
\eqref{eq:KBc_ee} and $\Ke=K+\veps$ as
\begin{align}
\calE_\ee[G(K)]&=\int\!Bc G(K_\ee) c\frac{\Ke^2}{K_\ee G(K_\ee)}
c G(K_\ee)c\frac{\Ke^2}{K_\ee G(K_\ee)} ,
\\
\calF_\ee[G(K)]&
=\int\!BcG(K_\ee)c\Ke^2\CR{\frac{1}{K_\ee G(K_\ee)}}{c}
\Ke^2\CR{G(K_\ee)}{c}\frac{\Ke^2}{K_\ee G(K_\ee)} .
\end{align}
The $\EOMtest$ can be non-trivial if $\calE_\ee$ is of $O(1/\veps)$
and/or $\calF_\ee$ is of $O(1/\eta)$.

In \cite{HK}, we considered only the EOM breaking by $\veps$, namely,
$\veps\times\calE_{\veps}$ with $\calE_\veps=\calE_{\veps,\eta=0}$
given by \eqref{eq:calE_e}.
We may simplify \eqref{eq:Eomtest=eE+etF} by putting $\eta=0$
($\veps=0$) in $\calE_\ee$ (in $\calF_\ee$)
to consider $\veps\times\calE_\veps+\eta\times\calF_{\veps=0,\eta}$.
Then, we can easily show by using the property
\eqref{eq:invBcccc} that $\calE_\veps$ and
$\calF_{\veps=0,\eta}$ have a simple relationship:
\begin{equation}
\calF_{\veps=0,\eta}[G(K)]=\calE_\eta[G(1/K)] ,
\end{equation}
where $\calE_\eta$ is \eqref{eq:calE_e} with $\veps$ replaced with
$\eta$.
Therefore, the $\EOMtest$ is expressed only in terms of
$\calE_\veps$:
\begin{equation}
\EOMtest[\Psi]=\lim_{\veps\to +0}\veps\times\Bigl\{
\calE_\veps[G(K)]+\calE_\veps[G(1/K)]\Bigr\} .
\label{eq:EOMtest=E+E}
\end{equation}
This is also understood without explicit calculations by applying the
invariance \eqref{eq:invBcccc} to $\EOMtest$
\eqref{eq:Eomtest=eE+etF}, and using the facts that the inversion
commutes with $\QB$, $\CR{\Mt_{1/K}}{\QB}=0$, and that the inversion
of $\Psi_\ee$ \eqref{eq:Psi_ee} is given by
\begin{equation}
\wt{\Psi_\ee}
=\Psi\bigr|_{(K,B,c)\to(\wt{K_\ee},\wt{B_\ee},\wt{c_\ee})}
=c\frac{\bigl(K+\eta(1+\veps K)\bigr)^2}{K'_\etep G(1/K'_\etep)}
Bc\left(1-G(1/K'_\etep)\right) ,
\label{eq:wtPsi_ee}
\end{equation}
where $(\wt{K_\ee},\wt{B_\ee},\wt{c_\ee})$ and $K'_\etep$ are defined
in \eqref{eq:wtKBC_ee} and \eqref{eq:KBc'_ee}, respectively.
Eq.\ \eqref{eq:wtPsi_ee} should be compared with \eqref{eq:Psi_ee}.

Eq.\ \eqref{eq:EOMtest=E+E} implies that, for $\Psi$ specified by
$G(K)$ with the behavior \eqref{eq:G(K)atK=0K=infty}, its EOM in the
strong sense is give by
\begin{equation}
\EOMtest[\Psi]=\eomT(n_0)+\eomT(n_\infty) ,
\label{eq:EOMtest=B+B}
\end{equation}
where $\eomT(n)$ is the function appearing in \eqref{eq:EOM1F1}.

\subsection{Solutions with $\bm{\calN=\pm 2}$}
\label{subsec:solwithN=pm2}

Now from our results \eqref{eq:calN=-n_0+n_infty+A} and
\eqref{eq:EOMtest=B+B} including contributions from both the
eigenvalues $K=0$ and $K=\infty$, we see that there exist satisfactory
solutions with integer $\calN$ and satisfying the EOM in the strong
sense only when both $n_0$ and $n_\infty$ are either of $0$ and
$\pm 1$. This implies that we can extend the construction of
satisfactory solutions to the cases $\calN=\pm 2$ (i.e., three branes
and ``$(-1)$ brane''). The corresponding $G(K)$ are, for example,
\begin{equation}
G(K)=\frac{(1+K)^2}{K}\quad(\calN=2),\qquad
G(K)=\frac{K}{(1+K)^2}\quad(\calN=-2) .
\label{eq:G_N=pm2}
\end{equation}
We can also construct satisfactory solutions with $\calN=\pm 1$ using
$K=\infty$. They are, for example,
\begin{equation}
G(K)=1+K\quad (\calN=1),\qquad
G(K)=\frac{1}{1+K}\quad (\calN=-1) .
\label{eq:G_N=pm1}
\end{equation}
Note that $G(K)=K$ has $(n_0,n_\infty)=(1,-1)$ and represents the
perturbative vacuum with $\calN=0$.

Eqs.\ \eqref{eq:calN=-n_0+n_infty+A} and \eqref{eq:EOMtest=B+B} shows
that satisfactory solutions with $\abs{\calN}\ge 3$ are impossible.
One might think that solutions with $\abs{\calN}\ge 3$ can be
constructed by adopting $G(K)$ which has zeros/poles in $0<K<\infty$
in addition to $K=0$ and/or $K=\infty$.
However, recall that we rely on the Schwinger parameter representation
of quantities containing $G(K)$ or $1/G(K)$ in evaluating $\calN$.
For a quantity having zeros/poles at finite $K>0$, its Schwinger
parameter representation does not exist; for example,
$1/(K-a)$ cannot be expressed as
$\int_0^\infty\!dt\,e^{-(K-a)t}$ for $K<a$.
We will return to this problem in the next section.

\subsection{Direct evaluation of $\calN$}
\label{subsec:Direct}

In the above arguments, we treated the singularity at $K=\infty$ by
mapping it to $K=0$ through the inversion \eqref{eq:Kto1/K}.
In this subsection, we present the evaluation of $\calN$ by treating
$K=\infty$ directly in the regularization \eqref{eq:KBc_ee}.
Here, we take
\begin{equation}
G(K)=1+K ,
\label{eq:G=1+K}
\end{equation}
as an example.

Let us summarize the formulas for $\calN$ \eqref{eq:intQBA} in the
regularization \eqref{eq:KBc_ee}. The winding number of a pure-gauge
solution $\Psi=U\QB U^{-1}$ is given by
\begin{equation}
\calN=\int\left(\QB\calA\right)_{(K,B,c)\to(K_\ee,B_\ee,c_\ee)}
=\pi^2\int_0^1\!d\tau\,\calB(\tau) ,
\label{eq:calN=intcalB}
\end{equation}
in terms of the function $\calB(\tau)$ which vanishes without
regularization:
\begin{equation}
\calB(\tau)=\int\!\left[\QB\left(
\Psi_\tau *\deriv{\Psi_\tau}{\tau}\right)
\right]_{(K,B,c)\to(K_\ee,B_\ee,c_\ee)}
=\int\!\left(\deriv{\Psi_\tau}{\tau}*\Psi_\tau *\Psi_\tau\right)_{
(K,B,c)\to(K_\ee,B_\ee,c_\ee)} .
\label{eq:calB=intQB...}
\end{equation}
Here, $\Psi_\tau$ with a parameter $\tau$ ($0\le\tau\le 1$) is given
by $\Psi_\tau=U_\tau\QB U_\tau^{-1}$ in terms of
$U_\tau=1-Bc\left(1-G_\tau(K)\right)$ which interpolates
$U_{\tau=0}=1$ and $U_{\tau=1}=U$ (hence $G_{\tau=0}(K)=1$ and
$G_{\tau=1}(K)=G(K)$).
Explicitly, $\calB(\tau)$ is given by
\begin{equation}
\calB(\tau)=\int\!Bc G_\tau c\frac{\Ke^2}{K_\ee}
\CR{\frac{1}{G_\tau}\deriv{G_\tau}{\tau}}{c}\frac{\Ke^2}{K_\ee}
\CR{\frac{1}{G_\tau}}{c}\frac{\Ke^2}{K_\ee} ,
\label{eq:calB=intBc...}
\end{equation}
with $G_\tau=G_\tau(K_\ee)$.

As $G_\tau(K)$ for $G(K)$ of \eqref{eq:G=1+K}, we adopt
$G_\tau(K)=1+\tau K$. For the present $G(K)$ we are allowed to
put $\veps=0$ since the integrand of \eqref{eq:calB=intBc...} is
regular at $K=0$. In this simplification, $\Ke^2/K_\ee$ in
\eqref{eq:calB=intBc...} is reduced to $K(1+\eta K)$, and we have
\begin{equation}
G_\tau(K_{\veps=0,\eta})=1+\frac{\tau K}{1+\eta K} .
\label{eq:G_tau=1+tauK/(1+etaK)}
\end{equation}
Let us first consider evaluating $\calB(\tau)$ \eqref{eq:calB=intBc...}
by using the $(s,z)$-integration formula \eqref{eq:szint} for the
correlator. Adding a large semi-circle in $\Re z<0$ to closed the
contour of the $z$-integration, the poles in the $z$-plane are located
at three points; $z=-1/(\tau+\eta)$ and $-1/(\tau+\eta)\pm 2\pi i/s$.
Then, carrying out the $s$-integration, we obtain the desired result:
\begin{equation}
\pi^2\calB(\tau)=\frac{3\eta\tau^2}{(\tau+\eta)^4}
\underset{\eta\to 0}{\to}\delta(\tau) .
\label{eq:pi^calB=delta}
\end{equation}

Next, we consider another way of calculating $\calB(\tau)$.
Note that $\calB(\tau)$ is given as a sum of terms of the form
$\int\!Bc F_1(K)cF_2(K)cF_3(K)cF_4(K)$.
We carry out the integrations over the Schwinger parameters $t_i$ for
$F_i(K)$ without using the $(s,z)$-trick of inserting
$1=\int_0^\infty\!ds\,\delta(s-\sum_i t_i)$.
In this case, there arises a delicate problem.
In order to get the same result as \eqref{eq:pi^calB=delta},
we need a special care for $F_i(K)$ ($i=1,2,3$) which are sandwiched
between two $c$'s;
we must use the partial fraction decomposed form without any $K$ in
the denominators for each $F_i(K)$ ($i=1,2,3$).
For example, for $F_1(K)=G_\tau(K_{0,\eta})$ of
\eqref{eq:G_tau=1+tauK/(1+etaK)}, we must use, instead of
$c\,G_\tau(K_{0,\eta}) c=c \tau K/(1+\eta K) c$,
the following expression:
$$
c\,G_\tau(K_{0,\eta}) c=-\frac{\tau}{\eta}c\frac{1}{1+\eta K}c .
$$
The two $cG_\tau(K_{0,\eta}) c$'s should be the same thing if we can
use $c^2=0$, which correspond to $t_1=0$. The problem arises when
other $t_i$'s are also infinitesimal and the ratio $t_1/t_i$ is
finite.

To explain the problem in more detail, let us consider a simpler
example:
\begin{equation}
\int\!Bc\frac{1}{1+K}cK^p c K^qc\frac{1}{a+K}
=\int_0^\infty\!dt_1\,e^{-t_1}\int_0^\infty\!dt_4\,e^{-at_4}
f(t_1,t_4),
\label{eq:exone}
\end{equation}
and
\begin{equation}
\int\!Bc\frac{-K}{1+K}cK^p c K^qc\frac{1}{a+K}
=\int_0^\infty\!dt_1\,e^{-t_1}\int_0^\infty\!dt_4\,e^{-at_4}
\Drv{}{t_1}f(t_1,t_4).
\label{eq:extwo}
\end{equation}
Carrying out the integration by parts with respect to $t_1$ in
\eqref{eq:extwo}, it is reduced to \eqref{eq:exone} if we can discard
the surface term. However, the surface term at $t_1=0$,
\begin{equation}
\lim_{t_1\to 0}\int_0^\infty\!dt_4\,e^{-at_4}\,f(t_1,t_4) ,
\label{eq:surfaceterm}
\end{equation}
is a subtle quantity.
We certainly have $f(t_1\to 0,t_4>0)\to 0$.
However, the scaling property
$f(\lambda t_1,\lambda t_4)=\lambda^{3-p-q}\,f(t_1,t_4)$
implies that the expansion of $f(t_1,t_4)$ in powers of $t_1$
starts with $t_1/t_4^{p+q-2}$. This series expansion does not make
sense for $p+q\ge 3$ since the $t_4$-integration is divergent at
$t_4=0$.

If we use the $(s,z)$-integration formula \eqref{eq:szint}, there
is no problem of $c^2\ne 0$.
This is because, in the derivation of \eqref{eq:szint} by inserting
$1=\int_0^\infty\!ds\,\delta(s-\sum_it_i)$, the quantity
$t_i/\sum_{j=1}^4t_j$ in the original expression is replaced with
$t_i/s$, and the $t_i$-integrations are carried out with $s$ kept
fixed. Therefore, there appear no subtle ratios $t_i/t_j$.
That $c^2=0$ is ensured in \eqref{eq:szint} is also seen from the fact
$\calG(F_1,F_2,F_3,F_4)$ \eqref{eq:calG} vanishes when at least
one of $F_{1,2,3}$ is a constant.

\section{A proposal for solutions with $\bm{\abs{\calN}\ge 3}$}
\label{sec:calNge3}

We found in the previous section that satisfactory multi-brane
solutions with $\abs{\calN}\ge 3$ cannot be constructed by using only
the singularities at $K=0$ and $\infty$.
One might be tempted to adopt $G(K)$ which has zeros/poles in
$0<K<\infty$. However, such $G(K)$ is problematic since quantities
containing $G(K)$ or $1/G(K)$ do not have well-defined Schwinger
parameter representations,
as we stated in Sec.\ \ref{subsec:solwithN=pm2}.
In this section, we propose a way of constructing satisfactory
solutions with $\abs{\calN}\ge 3$ by employing such apparently
``dangerous'' $G(K)$.

To explain our proposal, let us consider the following $G(K)$:
\begin{equation}
G(K)=\frac{(K+1)^3}{K(K-a)}\qquad (a>0).
\label{eq:G(K)ex}
\end{equation}
Our idea of defining the winding number and the $\EOMtest$ for $\Psi$
specified by this $G(K)$ is as follows. First, we divide the original
eigenvalue space of $K$, $[0,\infty]$, into two intervals, $[0,a]$ and
$[a,\infty]$.
Then, we expand each of the two intervals to $[0,\infty]$ by a linear
fractional transformation with real coefficients,
$K\mapsto(\alpha K+\beta)/(\gamma K+\delta)$, which is a real-to-real
and one-to-one mapping.
For example, we take the transformations
$K\mapsto g_1(K)=a K/(1+K)$ and $K\mapsto g_2(K)=a+K$
for the first and the second intervals, respectively.

After this preparation, we define $\calN$ and $\EOMtest$ for $\Psi$
specified by $G(K)$ \eqref{eq:G(K)ex} by the sum of those for
$\Mt_{g_{1,2}}(\Psi)$, the \Mn-transform \eqref{eq:Mtransf} of $\Psi$
associated with the maps $g_{1,2}(K)$:
\begin{equation}
\calO[\Psi]=\calO[\calM_{g_1}(\Psi)]+\calO[\calM_{g_2}(\Psi)],
\qquad
\left(\calO=\calN,\ \EOMtest\right) .
\label{eq:O=O+O}
\end{equation}
Of course, each of the two terms on the RHS of \eqref{eq:O=O+O} should
be evaluated by using the regularization of Sec.\ \ref{subsec:Reg}.
Recall that, for $\Psi$ of the form \eqref{eq:Psi=}, $\calM_g(\Psi)$ is
again given by \eqref{eq:Psi=} with $G(K)$ replaced with
$G\bigl(g(K)\bigr)$.
Now, since $G\bigl(g_{1,2}(K)\bigr)$ has no zeros/poles in
$0<K<\infty$, we can apply the arguments of Sec.\ 2 to
$G\bigl(g_{1,2}(K)\bigr)$.
The RHS of \eqref{eq:O=O+O} is determined only by $(n_0,n_\infty)$
of $G(g_{1,2}(K))$ (see eq.\ \eqref{eq:G(K)atK=0K=infty});
$(n_0,n_\infty)=(-1,-1)$ for both $G\bigl(g_{1,2}(K)\bigr)$.
This implies that $\Psi$ of $G(K)$ \eqref{eq:G(K)ex} is a satisfactory
solution representing five branes with $\calN=4$ and $\EOMtest=0$.

Generalization of the above example is manifest.
Suppose that $G(K)$ has zeros/poles at $K=a_i$ ($i=0,1,\cdots,m+1$)
with $a_0=0$ and $a_{m+1}=\infty$, and the leading behavior of $G(K)$
near $K=a_i$ is $G(K)\sim (K-a_i)^{n_i}$ $(i=0,\cdots,m)$.
For this $G(K)$, we divide the eigenvalue space $[0,\infty]$ of $K$
into $m+1$ intervals $[a_{i-1},a_i]$ ($i=1,\cdots,m+1$), and expand
each of the intervals to $[0,\infty]$ by the map $K\mapsto g_i(K)$
with $g_i(K)$ given, for example, by
\begin{equation}
g_i(K)=a_{i-1}+\frac{(a_i-a_{i-1}) K}{1+K}\quad
(i=1,\cdots, m),\qquad
g_{m+1}(K)=a_m+K .
\end{equation}
Then, we define $\calO[\Psi]=\calN[\Psi]$ and $\EOMtest[\Psi]$ for
$\Psi$ specified by the present $G(K)$ by
\begin{equation}
\calO[\Psi]=\sum_{i=1}^{m+1}\calO[\Mt_{g_i}(\Psi)] .
\label{eq:O=sumO}
\end{equation}
Concretely, we have
\begin{align}
\calN[\Psi]&=-n_0-n_\infty+\Anom(n_0)+A(n_\infty)
+2\sum_{i=1}^m\left[-n_i+\Anom(n_i)\right] ,
\label{eq:calNgen}
\\
\EOMtest[\Psi]&=\eomT(n_0)+\eomT(n_\infty)+2\sum_{i=1}^m \eomT(n_i) ,
\label{eq:EOMtestgen}
\end{align}
which is a generalization of \eqref{eq:calN=-n_0+n_infty+A} and
\eqref{eq:EOMtest=B+B}.
Note that the contribution from $a_i$ ($i=1,\cdots,m$) is multiplied
by $2$. This is because $a_i$ is a boundary of the two intervals
$[a_{i-1},a_i]$ and $[a_i,a_{i+1}]$.
{}From our results \eqref{eq:calNgen} and \eqref{eq:EOMtestgen}, we
find that satisfactory multi-brane solutions can be constructed for
any integer $\calN$ by using $G(K)$ which has simple zeros/poles in
$0<K<\infty$ in addition to those at $K=0$ and $\infty$.

Let us examine the consistency of our definition \eqref{eq:O=sumO}.
First, consider $G(K)$ which has no zeros/poles in $0<K<\infty$
and hence needs no division into intervals. For this $G(K)$, the LHS of
\eqref{eq:O=sumO} is already given in Sec.\ 2, and it agrees with the
RHS defined by arbitrarily chosen $a_i$.
This is because the present linear fractional transformations $g_i(K)$
never produce new zeros/poles from $G(K)$.
Second, the functions $g_i(K)$ are not uniquely determined by $a_i$
alone. For example, we could have chosen $g_1(K)=a/(1+\alpha K)$ and
$g_2(K)=a+\beta K$ ($\alpha,\beta>0$) for $G(K)$ of
\eqref{eq:G(K)ex}.
However, $\calN$ and $\EOMtest$ defined by \eqref{eq:O=sumO}
are not affected by the arbitrariness of $g_i(K)$ since
$\calO[\Mt_{g_i}(\Psi)]$ is determined only by $(n_{i-1},n_i)$.
Finally, we considered here only $\calN$ and $\EOMtest$ as
$\calO$ in \eqref{eq:O=sumO}. It is unknown whether possible other
observables associated with $\Psi$ are also ``topological'' ones
determined only by $n_i$. This is an important question to be
clarified.

In the above arguments, we have implicitly assumed that $G(K)$ has no
zeros/poles with $\Im K\ne 0$ and $\Re K>0$.
Such $G(K)$ is also dangerous and needs division into intervals.
For example, if $G(K)$ has zeros/poles at $K=a\pm i b$ ($a,b>0$),
we need two intervals $[0,a]$ and $[a,\infty]$.
However, these complex zeros/poles cannot contribute to $\calN$ and
$\EOMtest$ defined by \eqref{eq:O=sumO} since $G(K)$ has no
zeros/poles at $K=a$.

\section{Conclusions}
\label{sec:Conclusions}

In this paper, we presented a new way of constructing satisfactory
multi-brane solutions in CSFT, namely, solutions carrying integer
winding number $\calN$ and satisfying the EOM in the strong sense.
We found in the previous paper \cite{HK} that the origin of the
non-zero value of the topological quantity $\calN$ is the singularity
at $K=0$, and that stronger singularity leads to non-integer $\calN$
and the breaking of the EOM in the strong sense. Therefore, only
satisfactory solutions with $\calN=0, \pm 1$ were possible.
Our new construction consists of two steps, and, in both steps, the
\Mn\ transformation \eqref{eq:Mtransf} which keeps the $\KBc$ algebra
plays important roles.
The first step is based on our finding \eqref{eq:invBcccc} that the
correlator is invariant under the inversion, the \Mn\ transformation
which interchanges $K$ with $1/K$. This property implies that the
singularities at $K=\infty$ as well as at $K=0$ can be the origin
of $\calN$, and allows us to extend the construction of satisfactory
multi-brane solutions to the cases $\calN=\pm 2$.
In our second step, we further extended
the construction to the cases $\abs{\calN}\ge 3$.
For this, we took $G(K)$ which has zeros/poles in $0<K<\infty$
besides at $K=0,\infty$ and hence does not give well-defined $\calN$.
We proposed a way of making $\calN$ and $\EOMtest$ for this $G(K)$
well-defined by dividing the eigenvalue space of $K$, $[0,\infty]$,
into a sum of safe intervals, and mapping each interval to
$[0,\infty]$ by suitable \Mn\ transformations.

By our construction presented in this paper, satisfactory solutions
with any integer $\calN$ is now possible. Among them, the solutions
with $\abs{\calN}\le 2$ which are based on singularities at $K=0$
and/or $K=\infty$ are firm.
On the other hand, for our proposal of the solutions with
$\abs{\calN}\ge 3$, further study is necessary to conform that it is a
fully consistent one.
In particular, it is a problem whether all the physical quantities
related to the solution can be well-definedly given by
\eqref{eq:O=sumO}. For this, it would be necessary that all the
observables in CSFT are ``topological'' ones.

Besides the problem of solutions with $\abs{\calN}\ge 3$, there are
many questions to be clarified concerning our construction.
For example, there have been analyses of multi-brane solutions by studying
the gauge invariant observables \cite{MS2,BI} and the boundary states
\cite{Takahashi,MNT}.
It is interesting to examine how the $K=\infty$ eigenvalue can affect
these analyses.
Further study of the general property of \Mn\ transformation
\eqref{eq:Mtransf} is also necessary. In particular, we wish to know
whether the \Mn\ transformation has a simple CFT interpretation, and
whether there exists more general type of invariance of the correlator
under the \Mn\ transformation.

Finally, let us discuss in what sense the EOM should be satisfied by
the solutions $\Psi$, namely, for what class of test states $\Phi$ the
condition
\begin{equation}
\int\!\Phi*\left(
\QB\Psi_\ee+\Psi_\ee *\Psi_\ee\right)=0 ,
\label{eq:PhiEOM}
\end{equation}
should hold.
In this paper, we tried to construct solutions carrying an integer
$\calN$ and satisfying the EOM in the strong sense, i.e.,
\eqref{eq:PhiEOM} with $\Phi=\Psi_\ee$, by calling them
``satisfactory solutions''.
If the EOM in the strong sense holds, the winding number $\calN$ is
directory related to the energy density.
However, it was shown, within the framework of the
$\Ke$-regularization, that \eqref{eq:PhiEOM} is violated for $\Psi$
corresponding to $G(K)$ singular at $K=0$ when we take as the test
state $\Phi$ the Fock vacuum \cite{MS2}.\footnote{
For $\Psi$ corresponding to $G(K)=1+K$, the EOM condition
\eqref{eq:PhiEOM} holds against the Fock vacuum in our
regularization for $K=\infty$. This is consistent with the result
of \cite{Masuda} in a different regularization.
However, we think this fact not so important for our construction of
multi-brane solutions since the singularities both at $K=0$ and
$K=\infty$ are necessary.
}
This is the case also in our $(K_\ee,B_\ee,c_\ee)$-regularization 
\eqref{eq:KBc_ee} introduced here since $\eta$ has no effect
on the singularity at $K=0$.
Indeed, the Fock states are ``natural states'' for the perturbative
vacuum, namely, they are states representing the excitations around
a single D-brane.
However, is it necessary to demand \eqref{eq:PhiEOM} against $\Phi$
in the Fock space which have no special meanings for multi-brane
solutions?
Furthermore, it seems impossible to demand \eqref{eq:PhiEOM} for an
arbitrary $\Phi$; for any given $\Psi$ we could make \eqref{eq:PhiEOM}
non-vanishing by taking a test state $\Phi$ which carries strong
enough singularities at $K=0$ and/or $K=\infty$.

To consider for what class of $\Phi$ the EOM condition
\eqref{eq:PhiEOM} should be satisfied, let us recall the process of
obtaining the action of the fluctuation $\Phi$ around a multi-brane
solution $\Psi$:
\begin{equation}
S[\Psi_\ee+\Phi]=S[\Psi_\ee]
+\int\!\left(\frac12\Phi*Q_\Psi \Phi+\frac13\Phi^3\right),
\end{equation}
where $Q_\Psi$ is the BRST operator around the solution $\Psi$,
and we have used the EOM condition \eqref{eq:PhiEOM} to drop the term
linear in $\Phi$. Then, obviously \eqref{eq:PhiEOM} must hold against
$\Phi$ representing the open string excitations specified by
$Q_\Psi$.\footnote{
See \cite{EMRG} for an analysis of lump solutions by demanding the EOM
only against the space of fluctuations around the solution.
}
It is our future problem to carry out this analysis and
clarify the relationship to the EOM in the strong sense.
On the other hand, the Fock states would be far beyond the space of
finite mass excitations. We do not know whether we have to demand
\eqref{eq:PhiEOM} for such $\Phi$ outside the space of the
perturbative fluctuation modes around $\Psi$.

The analysis of the fluctuation modes around the solutions is itself
an important problem. For a solution with $\calN\ge 1$, we must
show that there is the $(\calN+1)^2$ degeneracy of the open string
excitations on $\calN+1$ branes. For solutions with $\calN\le -2$, we
must clarify whether such ``ghost branes'' can really exist, and if
so, what their physical meanings are, through the analysis of the
fluctuations.

\section*{Acknowledgments}
We would like to thank Isao Kishimoto, Toru Masuda, Toshifumi Noumi,
Yuji Okawa, Martin Schnabl, Daisuke Takahashi and Tomohiko Takahashi
for valuable discussions.
The work of H.~H.\ was supported in part by a Grant-in-Aid for
Scientific Research (C) No.~21540264 from JSPS.
The work of T.~K.\ was supported in part by a Grant-in-Aid for
JSPS Fellows No.~24$\cdot$1601.

\appendix

\section{Derivation of eqs.\ (\ref{eq:calN=-n+Anom})
and  (\ref{eq:EOM1F1})}
\label{app:calNandEOMtest}

In this appendix, we present the derivation of eq.\
\eqref{eq:calN=-n+Anom} for the winding number $\calN$ and
eq.\ \eqref{eq:EOM1F1} for the EOM in the strong sense by taking into
account only the zeros/poles of $G(K)$ at $K=0$.

First, let us consider $\calN$ \eqref{eq:calN=intcalB} given in terms
of $\calB(\tau)$ \eqref{eq:calB=intQB...}.
Since only the singularity at $K=0$ is of significance, we may start
with \eqref{eq:calB=intBc...} with $\eta=0$, which has the following
convenient expression:
\begin{align}
\calB(\tau)&=\int\!\QB\!\left\{Bc\,G_\tau(\Ke)c\Ke\!
\CR{\frac{(d/d\tau)G_\tau(\Ke)}{G_\tau(\Ke)}}{c}
\frac{\Ke}{G_\tau(\Ke)}\right\}
\nn\\
&\qquad
+\veps\int\!c\,G_\tau(\Ke)c\Ke\!
\CR{\frac{(d/d\tau)G_\tau(\Ke)}{G_\tau(\Ke)}}{c}
\frac{\Ke}{G_\tau(\Ke)} .
\label{eq:calB=intQB+vepsint}
\end{align}
We can drop the first term since it is the integration of the
BRST-transform of a quantity which is perfectly regular for $\veps>0$.
The second term is multiplied by $\veps$ and therefore can be
non-trivial only if the integration gives a $O(1/\veps)$ quantity.
Therefore, we have only to take the leading term of $G(K)$ near
$K=0$, which, as we will see, gives the desired $O(1/\veps)$ term.
Here, we consider the case that $G(K)$ has a pole at $K=0$:
\begin{equation}
G(K)\sim K^{-m}\quad (K\sim 0,\ m\ge 1) .
\label{eq:GsimK^-m}
\end{equation}
The treatment of the case $m<0$ is quite similar.

We take the parameter $\tau$ in such a way that $\Psi_\tau$
($0\le\tau<1$) and $\Psi_{\tau=1}$ represent the perturbative vacuum
and the non-trivial one $\Psi$, respectively.
Then, $G_\tau(K)$ must have no zeros/poles at $K=0$ for $\tau<1$.
Therefore, as the leading term of $G_\tau(K)$, we can take without
loss of generality
\begin{equation}
G_\tau(K)\sim\left(1-\tau +K\right)^{-m} .
\label{eq:Gtausim1-tau+K^-m}
\end{equation}
Substituting this into the second term of
\eqref{eq:calB=intQB+vepsint}, $\calB(\tau)$ is now given by
\begin{align}
\calB(\tau)&=-m\veps\biggl\{
\Dtau\int\!c\frac{1}{(\Ke+\Dtau)^m} c\frac{1}{\Ke+\Dtau}c
\Ke(\Ke+\Dtau)^m
\nn\\
&\qquad\qquad\qquad
+\int\!c\frac{1}{(\Ke+\Dtau)^m} cKc\Ke(\Ke+\Dtau)^{m-1}
\biggr\} ,
\end{align}
with $\Dtau=1-\tau$. We can evaluate this $\calB(\tau)$ by using the
$(s,z)$-integration formula \eqref{eq:szint} with $F_4=1$.
First, we carry out the $z$-integration by adding a large semi-circle
in $\Re z<0$ to close the contour and evaluating residues at
$z=-\Dtau-\veps$ and $z=-\Dtau-\veps\pm 2\pi i/s$.
Then, the $s$-integration is elementary and gives
\begin{align}
\pi^2\calB(\tau)&=\frac{m}{2}\sum_\pm\Biggl\{
d_1(\tau)+\frac12\,d_2(\tau)\sum_{k=0}^{m-1}\binom{m}{m-k-1}
p^{(\pm)}_{k,1}
\nn\\
&\quad
+\frac12\,d_3(\tau)\left[
\sum_{k=0}^{m-1}\left(\binom{m}{m-k-1}-\binom{m-1}{m-k-1}\right)
p^{(\pm)}_{k,0}
-\sum_{k=0}^m\binom{m+1}{m-k}p^{(\pm)}_{k,0}\right]
\nn\\
&\quad
+d_4(\tau)\left[
\sum_{k=0}^m\binom{m}{m-k}p^{(\pm)}_{k,-1}
-\sum_{k=0}^{m-1}\binom{m-1}{m-k-1}p^{(\pm)}_{k,-1}\right]
\Biggr\} ,
\end{align}
where $p^{(\pm)}_{k,\ell}$ and $d_{1,2,3,4}(\tau)$ are defined by
\begin{equation}
p_{k,\ell}^{(\pm)}=\frac{(\pm 2\pi i)^{k+\ell}}{k!} ,
\end{equation}
and
\begin{equation}
d_1(\tau)=\frac{2}{\veps(w+1)^3},\quad
d_2(\tau)=\frac{1}{\veps(w+1)^2},\quad
d_3(\tau)=\frac{2w}{\veps(w+1)^3},\quad
d_4(\tau)=\frac{3w^2}{\veps(w+1)^4},\quad
\label{eq:d}
\end{equation}
with $w=\Dtau/\veps$.
The four functions in \eqref{eq:d} are all reduced in the limit
$\veps\to +0$ to the delta function $\delta(1-\tau)$ satisfying
$\delta(1-\tau)=0$ for $\tau<1$ and
$\int_0^1\!d\tau\,\delta(1-\tau)=1$.
Using this fact and the definition of the confluent hypergeometric
series,
\begin{equation}
{}_1F_1(\alpha,\gamma;z)=\sum_{k=0}^\infty
\frac{\alpha(\alpha-1)\cdots(\alpha-k+1)}{
\gamma(\gamma-1)\cdots(\gamma-k+1)}\frac{z^k}{k!},
\label{eq:1F1}
\end{equation}
we find that
\begin{equation}
\pi^2\calB(\tau)\underset{\veps\to 0}{\to}
\bigl(m+\Anom(-m)\bigr)\,\delta(1-\tau) ,
\end{equation}
with $\Anom(n)$ given by \eqref{eq:Anom}.

Next, let us calculate $\EOMtest$ \eqref{eq:EOMtest}.
Since $\calE_\veps$ \eqref{eq:calE_e} is also multiplied by $\veps$,
we have only to take the leading term \eqref{eq:GsimK^-m} of $G(K)$ as
in the case of the second term of \eqref{eq:calB=intQB+vepsint}:
\begin{equation}
\calE_\veps=\int\!Bc\frac{1}{\Ke^m}c\Ke^{m+1}c\frac{1}{\Ke^m}
c\Ke^{m+1} .
\end{equation}
Again using the $(s,z)$-integration formula \eqref{eq:szint}, we
obtain $\calE_\veps=\eomT(-m)/\veps$ with $\eomT(n)$ given by
\eqref{eq:EOM1F1}.

\section{Proof of eq.\ (\ref{eq:invBcccc})}
\label{app:Proof_invBcccc}

In this appendix, we present a proof of the marvelous property
\eqref{eq:invBcccc} of the correlator.
Since any $\int\!\calW(K,B,c)$ is reduced to the sum of terms of the
form $\int\!BcF_1(K)cF_2(K)cF_3(K)cF_4(K)$ by using \eqref{eq:CRalg},
it is sufficient to show \eqref{eq:invBcccc} for the latter.
Then, the RHS of \eqref{eq:invBcccc} is rewritten as follows:
\begin{align}
&\int\!\wt{B}\wt{c} F_1(\wt{K})\wt{c}F_2(\wt{K})\wt{c}
F_3(\wt{K})\wt{c}F_4(\wt{K})
=\int\!Bc F_1(\wt{K})cK^2Bc F_2(\wt{K})cK^2Bc F_3(\wt{K})cK^2Bc
F_4(\wt{K})
\nn\\
&=\int\!BcF_1(\wt{K})cK^2\CR{F_2(\wt{K})}{c}K^2\CR{F_3(\wt{K})}{c}
K^2 F_4(\wt{K}) ,
\label{eq:RHSinvBccc}
\end{align}
where we have used
\begin{equation}
Bcf_1(K) c f_2(K) Bc=\CR{f_1(K)}{c}f_2(K)Bc ,
\end{equation}
valid for any $f_{1,2}(K)$.

Our proof is based on the $(s,z)$-integration formula of the
correlator \cite{MS,MS2}, which reads
\begin{equation}
\int\!BcF_1(K)cF_2(K)cF_3(K)cF_4(K)
=\int_0^\infty\!ds\,\frac{s^2}{(2\pi)^3\,i}
\int_{-i\infty}^{i\infty}\!\frac{dz}{2\pi i}\,e^{sz}\,
\calG\bigl(F_1,F_2,F_3,F_4\bigr) ,
\label{eq:szint}
\end{equation}
where $\calG$ is given in our convention by
\begin{align}
&\calG\bigl(F_1,F_2,F_3,F_4\bigr)
=\Bigl[(\Ds F_1)F_2F_3'+F_1'F_2(\Ds F_3)
+\bigl(F_1(\Ds F_2)F_3\bigr)'
\nn\\
&\qquad
-\Ds(F_1F_2')\,F_3
-\Ds(F_1F_2)\,F_3'
-F_1'\Ds(F_2F_3)-F_1\Ds(F_2'F_3)
+\Ds(F_1F_2'F_3)\Bigr]F_4 ,
\label{eq:calG}
\end{align}
with $F_i=F_i(z)$, $F_i'=(d/dz)F_i(z)$ and
\begin{equation}
(\Ds F_i)(z)\equiv F_i\!\left(z-\frac{2\pi i}{s}\right)
-F_i\!\left(z+\frac{2\pi i}{s}\right) .
\end{equation}
Note that we are allowed to extend the range of the $s$-integration
in \eqref{eq:szint} to $(-\infty,\infty)$ since \eqref{eq:szint} has
been obtained by inserting
$\int_0^\infty\!ds\,\delta\!\left(s-\sum_i t_i\right)=1$
with $t_i\ge 0$ being the Schwinger parameter for $F_i(K)$.
In the rest of this proof, we use \eqref{eq:szint} with the
extended $s$-integration region.

It is sufficient to show \eqref{eq:invBcccc} for $F_i(K)=e^{-t_i K}$
($i=1,2,3,4$).
Then, \eqref{eq:RHSinvBccc} is given as the $(s,z)$-integration
\eqref{eq:szint} with $\calG$ replaced by
\begin{align}
&\calG\bigl(
e^{-t_1/z},z^2\,e^{-t_2/z},z^2\,e^{-t_3/z},z^2\,e^{-t_4/z}\bigr)
-\calG\bigl(e^{-t_1/z},z^2\,e^{-t_2/z},z^2,z^2\,e^{-(t_3+t_4)/z}\bigr)
\nn\\
&\qquad\qquad
-\calG\bigl(e^{-t_1/z},z^2,z^2\,e^{-(t_2+t_3)/z},z^2\,e^{-t_4/z}\bigr)
+\calG\bigl(e^{-t_1/z},z^2,z^2\,e^{-t_2/z},z^2\,e^{-(t_3+t_4)/z}\bigr).
\label{eq:G-G-G+G}
\end{align}
In order to relate the $(s,z)$-integration for \eqref{eq:RHSinvBccc}
to that for the LHS of \eqref{eq:invBcccc}, let us first make the
following change of integration variables in the former:
\begin{equation}
z\to\frac{1}{z},\qquad s\to z^2\,s .
\label{eq:changesz1}
\end{equation}
Note that this keeps $e^{sz}$ invariant.
Then, we obtain
\begin{equation}
\eqref{eq:RHSinvBccc}=\frac{1}{(2\pi)^3\,i}\int_{-\infty}^\infty
\!\!ds\int_{-i\infty}^{i\infty}\!\frac{dz}{2\pi i}\,
e^{\left(s-\sum_i t_i\right)z}\left(\calG_++\calG_-\right) ,
\label{eq:Gp+Gm}
\end{equation}
where $\calG_\pm$ is defined by
\begin{align}
\calG_\pm&=\pm s_\pm^2\biggl[
-t_3\left(e^{\pm(2\pi i/s_\pm)t_1}-1\right)
-t_1\left(e^{\pm(2\pi i/s_\pm)t_3}-1\right)
\nn\\
&\qquad\qquad
-(t_1+t_2+t_3)\left(e^{\pm(2\pi i/s_\pm)t_2}-1\right)
+(t_2+t_3)\left(e^{\pm(2\pi i/s_\pm)(t_1+t_2)}-1\right)
\nn\\
&\qquad\qquad
+(t_1+t_2)\left(e^{\pm(2\pi i/s_\pm)(t_2+t_3)}-1\right)
-t_2\left(e^{\pm(2\pi i/s_\pm)(t_1+t_2+t_3)}-1\right)
\biggr]\pm 4\pi^2t_1t_2t_3 ,
\label{eq:G_pm}
\end{align}
with $s_\pm=s\pm(2\pi i/z)$. Let us further make a change of variables
$s\to s\mp(2\pi i/z)$ in the part of integration \eqref{eq:Gp+Gm}
multiplied by $\calG_\pm$. This shift of $s$ is allowed since we have
$\calG_\pm=O(1/s^2)$ as $s\to\infty$ owing to the last term of
\eqref{eq:G_pm}.
Then, we finally obtain the following expression of
\eqref{eq:RHSinvBccc}:
\begin{align}
&\int_{-\infty}^\infty\!ds\,\frac{s^2}{(2\pi)^3\,i}
\int_{-i\infty}^{i\infty}\!\frac{dz}{2\pi i}\,
e^{\left(s-\sum_i t_i\right)z}
\sum_\pm(\pm)\biggl\{-t_3\,e^{\pm(2\pi i/s)t_1}
-t_1\,e^{\pm(2\pi i/s)t_3}-t_2\,e^{\pm(2\pi i/s)(t_1+t_2+t_3)}
\nn\\
&\qquad
-(t_1+t_2+t_3)e^{\pm(2\pi i/s)t_2}
+(t_2+t_3)e^{\pm(2\pi i/s)(t_1+t_2)}
+(t_1+t_2)e^{\pm(2\pi i/s)(t_2+t_3)}
\biggr\} .
\label{eq:RHSinvBcccFinal}
\end{align}
We find that this is nothing but the LHS of \eqref{eq:invBcccc} with
$F_i(K)=e^{-t_i K}$. This ends a proof of \eqref{eq:invBcccc}.

Eq.\ \eqref{eq:invBcccc} can of course be confirmed by taking a
concrete $\calW(K,B,c)$ and calculating its both hand sides by using
the $(s,z)$-integration formula \eqref{eq:szint}.
If we carry out the calculation without using the $(s,z)$-trick, the
same care as we described in Sec.\ \ref{subsec:Direct} is
necessary.


\begin{thebibliography}{99}

\bibitem{CSFT}
  E.~Witten,
  ``Noncommutative Geometry and String Field Theory,''
  Nucl.\ Phys.\  B {\bf 268}, 253 (1986).

\bibitem{MS}
  M.~Murata and M.~Schnabl,
  ``On Multibrane Solutions in Open String Field Theory,''
  Prog.\ Theor.\ Phys.\ Suppl.\  {\bf 188}, 50 (2011)
  [arXiv:1103.1382 [hep-th]].

\bibitem{Takahashi}
  D.~Takahashi,
  ``The boundary state for a class of analytic solutions in open
  string field theory,''
  JHEP {\bf 1111}, 054 (2011) [arXiv:1110.1443 [hep-th]].

\bibitem{HK}
  H.~Hata and T.~Kojita,
  ``Winding Number in String Field Theory,''
  JHEP {\bf 1201}, 088 (2012) [arXiv:1111.2389 [hep-th]].

\bibitem{MS2}
  M.~Murata and M.~Schnabl,
  ``Multibrane Solutions in Open String Field Theory,''
  JHEP {\bf 1207}, 063 (2012)  [arXiv:1112.0591 [hep-th]].

\bibitem{EM}
  T.~Erler and C.~Maccaferri,
  ``Connecting Solutions in Open String Field Theory with Singular
  Gauge Transformations,''
  JHEP {\bf 1204}, 107 (2012) [arXiv:1201.5119 [hep-th]].

\bibitem{AldoArroyo}
  E.~Aldo Arroyo,
  ``Multibrane solutions in cubic superstring field theory,''
  JHEP {\bf 1206}, 157 (2012) [arXiv:1204.0213 [hep-th]].

\bibitem{MNT}
  T.~Masuda, T.~Noumi and D.~Takahashi,
  ``Constraints on a class of classical solutions in open string field
  theory,''
  arXiv:1207.6220 [hep-th].

\bibitem{BI}
  T.~Baba and N.~Ishibashi,
  ``Energy from the gauge invariant observables,''
  arXiv:1208.6206 [hep-th].

\bibitem{Okawa}
  Y.~Okawa,
  ``Comments on Schnabl's analytic solution for tachyon condensation in
  Witten's open string field theory,''
  JHEP {\bf 0604}, 055 (2006)
  [arXiv:hep-th/0603159].

\bibitem{ES}
  T.~Erler and M.~Schnabl,
  ``A Simple Analytic Solution for Tachyon Condensation,''
  JHEP {\bf 0910}, 066 (2009)
  [arXiv:0906.0979 [hep-th]].

\bibitem{Erler}
  T.~Erler,
  ``The Identity String Field and the Sliver Frame Level Expansion,''
  arXiv:1208.6287 [hep-th].

\bibitem{ErlerProc}
  T.~Erler,
  ``A simple analytic solution for tachyon condensation,''
  Theor.\ Math.\ Phys.\  {\bf 163}, 705 (2010)
  [Teor.\ Mat.\ Fiz.\  {\bf163}, 366 (2010)].

\bibitem{Masuda}
  T.~Masuda,
  ``Comments on new multiple-brane solutions based on Hata-Kojita
  duality in open string field theory,''
  arXiv:1211.2649 [hep-th].

\bibitem{EMRG}
  T.~Erler and C.~Maccaferri,
  ``Comments on Lumps from RG flows,''
  JHEP {\bf 1111}, 092 (2011)
  [arXiv:1105.6057 [hep-th]].

\end{thebibliography}
\end{document}